\documentclass[useAMS,usenatbib]{mn2e}
\usepackage[dvips]{graphicx}
\usepackage{amsmath}
\usepackage{amsfonts}
\usepackage{amssymb}

\title[Nightside Pollution of Exoplanet Transits]{Nightside Pollution of Exoplanet Transit Depths}
\author[Kipping \& Tinetti]{David M. Kipping$^{1,2}$\thanks{E-mail:
d.kipping@ucl.ac.uk}\footnotemark[1] \& Giovanna Tinetti$^{1}$ \\
$^{1}$Department of Physics and Astronomy, University College London, \\
       Gower Street, London WC1E 6BT, UK \\
$^{2}$Harvard-Smithsonian Center for Astrophysics \\
60, Garden St. Cambridge, MA, 02138, USA}
\begin{document}

\date{Accepted 2010 May 26. Received 2010 March 26; in original form 2010 March 26}

\pagerange{\pageref{firstpage}--\pageref{lastpage}} \pubyear{2010}

\maketitle

\label{firstpage}

\begin{abstract}
Out of the known transiting extrasolar planets, the majority are gas giants orbiting their host star at close proximity.  Both theoretical and observational studies support the hypothesis that such bodies emit significant amounts of flux relative to the host star, increasing towards infrared wavelengths.  For the dayside of the exoplanet, this phenomenon typically permits detectable secondary eclipses at such wavelengths, which may be used to infer atmospheric composition.  In this paper, we explore the effects of emission from the nightside of the exoplanet on the primary transit lightcurve, which is essentially a self-blend.  Allowing for nightside emission, an exoplanet's transit depth is no longer exclusively a function of the ratio-of-radii.  The nightside of an exoplanet is emitting flux and the contrast to the star's emission is of the order of $\sim 10^{-3}$ for hot-Jupiters.  Consequently, we show that the transit depth in the mid-infrared will be attenuated due to flux contribution from the nightside emission by $\sim 10^{-4}$.  We show how this effect can be compensated for in the case where exoplanet phase curves have been measured, in particular for HD 189733b.  For other systems, it may be possible to make a first-order correction by using temperature estimates of the planet.  Unless the effect is accounted for, transmission spectra will also be polluted by nightside emission and we estimate that a \emph{Spitzer} broadband spectrum on a bright target is altered at the 1-$\sigma$ level.  Using archived \emph{Spitzer} measurements, we show that the effect respectively increases the 8.0$\mu$m and 24.0$\mu$m transit depths by 1-$\sigma$ and 0.5-$\sigma$ per transit for HD 189733b. Consequently, we estimate that this would be $\sim$5-10$\sigma$ effect for near-future JWST observations.
\end{abstract}




\section{Introduction}

The field of exoplanetary science was galvanized by the discovery of the first transiting planet by \citet{cha00} and \citet{hen00}.  The transit discovery was made in the visible wavelength range and for the subsequent few years this was established as the normal practice in later observations and surveys (e.g. \citet{bro01}; \citet{bak04}; \citet{pol06}).  In response to the growing need for accurate parametrization of lightcurves, several authors produced equations modelling the lightcurve behavior including \citet{sac99}, \citet{sea03}, \citet{gim07} and \citet{kip08}.

In the last two years, the value of infrared measurements of transiting systems has become apparent with numerous pioneering detections; emission from a transiting planet \citep{dem05}, emission from a non-transiting planet \citep{har06}, an exoplanetary spectrum \citep{gri07}, detection of water in the atmosphere of HD 189733b \citep{tin07b}, methane \citep{swa08} and more recently carbon dioxide \citep{swa09}.  More details on the use of transmission spectroscopy as a tool for detecting molecular species can be found in \citet{sea00} and \citet{tin09}.  With JWST set to replace HST in the next decade, we can expect an abundance of high-precision infrared transits to be observed in order to detect more molecular species, perhaps including biosignatures \citep{sea05}.  In this paper, we discuss the consequences of significant nightside planetary emission on precise infrared transit lightcurves.  Nightside emission renders one of the original assumptions of transit theory invalid: `The planet may be treated as a black disc occulting the flux-emitting star'.  In the case of hot-Jupiter systems, the nightside of the planet is now also flux-emitting. This additional flux can be considered as a blend, but from the planet itself, i.e. a \emph{self-blend}.

Conceptually, it is very easy to see that this will cause mid-infrared transit depth measurements (and to a lesser degree in the visible and near-infrared range) to become underestimates of the true depth.  The reason is that there are two sources of flux, the star and the planet, and only one of these is being occulted, whilst the other is the blend source.  Note that we define the true transit depth as a purely geometric effect determined by the ratio-of-radii squared i.e. $d_{\mathrm{geo}} = (R_P/R_*)^2 = p^2$.  In this work, we will derive expressions estimating the amplitude of the effect, propose methods for correcting the lightcurves and apply them to two cases where the nightside emission of an exoplanet has been determined in the mid-infrared.

\section{Derivation}
\subsection{Depth dilution}

Let us define the total out-of-transit flux surrounding a transit event, as shown in figure 2a, to be given by:

\begin{equation}
F_{out,\mathrm{tra}} = F_{*} + F_{P,\mathrm{night}}
\end{equation}

where $F_*$ is the total flux received from the star and $F_{\textrm{P,night}}$ is the total flux received from the nightside of the planet, over a time interval of $\delta t$.  Let us assume that the star is a uniform emitter and that both the stellar and planetary total flux are invariable over the timescale of the transit event.  We may then write down the flux during a transit as a function of the ratio of the radii, $p$:

\begin{equation}
F_{in,\mathrm{tra}} = (1-p^2) \cdot  F_{*} + F_{P,\mathrm{night}}
\end{equation}

Note how the flux of the star has been attenuated as a result of the eclipse but the planetary flux is still present.  The observed transit depth in the flux domain, $d_{\mathrm{obs}}$, is defined by:

\begin{align}
d_{\mathrm{obs}} &= \frac{F_{out,\mathrm{tra}} - F_{in,\mathrm{tra}}}{F_{out,\mathrm{tra}}} \\
d_{\mathrm{obs}} &= \Big(\frac{F_*}{F_* + F_{P,\mathrm{night}}}\Big) \cdot d_{\mathrm{geo}}
\end{align}

Where $d_{\mathrm{geo}}$ is the geometric transit depth.  In the case of $F_{\textrm{P,night}} \rightarrow 0$, we recover the standard equation for the depth being $d_{\mathrm{obs}} = d_{\mathrm{geo}} = p^2$.  For cases where the nightside flux of the planet is non-negligible, the transit depth will therefore be affected.  We may re-write equation (4) as $d_{\mathrm{geo}} = B_{\mathrm{night}} \times d_{\mathrm{obs}}$ where we define:

\begin{equation}
B_{\mathrm{night}} = \frac{F_* + F_{P,\mathrm{night}}}{F_*}
\end{equation}

The blend source is at a much cooler temperature than the host star and thus the contrast between the two bodies is greater at infrared wavelengths. Consequently, the self-blending has a significant effect on infrared measurements (e.g. see \S5.1) but a much lower impact on visible wavelengths (e.g. see \S4.4). This makes the inclusion of such an effect paramount since visible and infrared measurements must be considered incommensurable unless this systematic is corrected for.

\subsection{The consequences for other parameters}

The largest effect of a blend is to underestimate the transit depth. However, we evaluate here the effect of the nightside blend, or indeed any kind of blend, on the other lightcurve parameters. For simplicity, we consider here a circular orbit and so we may use the expressions of \citet{sea03}. Assuming a blend factor given by $B$ we have $d_{obs} = p^2/B$. \citet{sea03} derived expressions for retrieving the impact parameter squared, $b^2$, and the semi-major axis (in stellar radii) squared, $(a/R_*)^2$, as a function of $t_T$, $t_F$ and $d$, where $t_T$ and $t_F$ are the $1^{\mathrm{st}}$-to-$4^{\mathrm{th}}$ and $2^{\mathrm{nd}}$-to-$3^{\mathrm{rd}}$ contact transit durations respectively. We may calculate the retrieved value of $b^2$ and $(a/R_*)^2$ in the case where a blend exists but we have negated it in our analysis.  Using equations (7) and (8) from \citet{sea03}, we find:

\begin{align}
b_{derived}^2 &= \frac{p^2 - (1 + p^2 -b^2) \sqrt{B} + B}{B} \nonumber \\
b_{derived}^2 &= b^2 + \frac{1}{2} (1-b^2 - p^2) (B-1) + \mathcal{O}[(B-1)^2]
\end{align}

\begin{align}
(a/R_*)_{derived}^2 &= (a/R_*)^2 \nonumber \\
\qquad& - \frac{(1-b^2-p^2) ((a/R_*)^2-(1+p)^2)}{2 ((1+p)^2-b^2)} (B-1) \nonumber \\
\qquad& + \mathcal{O}[(B-1)^2]
\end{align}

Equations (6) and (7) imply that negating the blending factor causes us to overestimate the impact parameter and underestimate $a/R_*$. It is important to recall that the lightcurve derived stellar density is found by taking the cube of $a/R_*$ and therefore will exacerbate any errors at this stage. We note that both equations give the expected results for $B=1$, i.e. no blend source present. 

These expressions have been derived assuming no limb darkening is present, which is typically a very good approximation for the wavelength range we are interested in. However, in reality the incorporation of limb darkening is easily implemented and demonstrated later in \S5. Therefore, equations (6) and (7) should not be used to attempt to correct parameters derived from fits not accounting for nightside pollution, rather they offer an approximate quantification of the direction and magnitude of the expected errors.

\section{Compensating for the Effect}
\subsection{Empirical method}

In the previous section we saw how the geometric transit depth, $d_{\mathrm{geo}}$, and the observed transit depth, $d_{\mathrm{obs}}$, are related by the factor $B_{\mathrm{night}}$.  Fortunately, $B_{\mathrm{night}}$ is an observable and can be obtained through measuring the phase curve of an extrasolar planet (for the first measured example, see \citet{knu07}).  With such a measurement, the difference between the day and nightside fluxes may be determined and thus $F_{P,\mathrm{night}}/F_*$ can be calculated.

Another possible method is to measure the secondary and primary transits without the intermediate phase curve information, which would require instruments with extremely stable calibration.  For a very inactive star, a highly calibrated instrument could, in principle, measure the nightside flux by measuring just the primary and secondary eclipses.  This would equate to an absolute calibration accurate to a fraction of the difference between the day and nightside fluxes, estimated to be $\sim 10^{-3}$ for HD 189733b, over one half of the orbital period.  We therefore estimate calibration requirements to be at least $\sim 10^{-4}$ between a $\sim 30$ hour period.  Whilst this would be a moot point for staring telescopes like \emph{Kepler} and \emph{CoRoT}, it is the infrared telescopes of \emph{Spitzer} and JWST that are most heavily affected by nightside pollution and these telescopes frequently slew around looking at different patches of the sky.  After the slewing we require the target to be at the same centroid position to within a fraction of a pixel.  Whether the whole phase curve or simply the eclipse-only observations are made, the same method may be used to correct both for the effects of nightside pollution.

\begin{figure*}
\begin{center}
\includegraphics[width=16.8 cm]{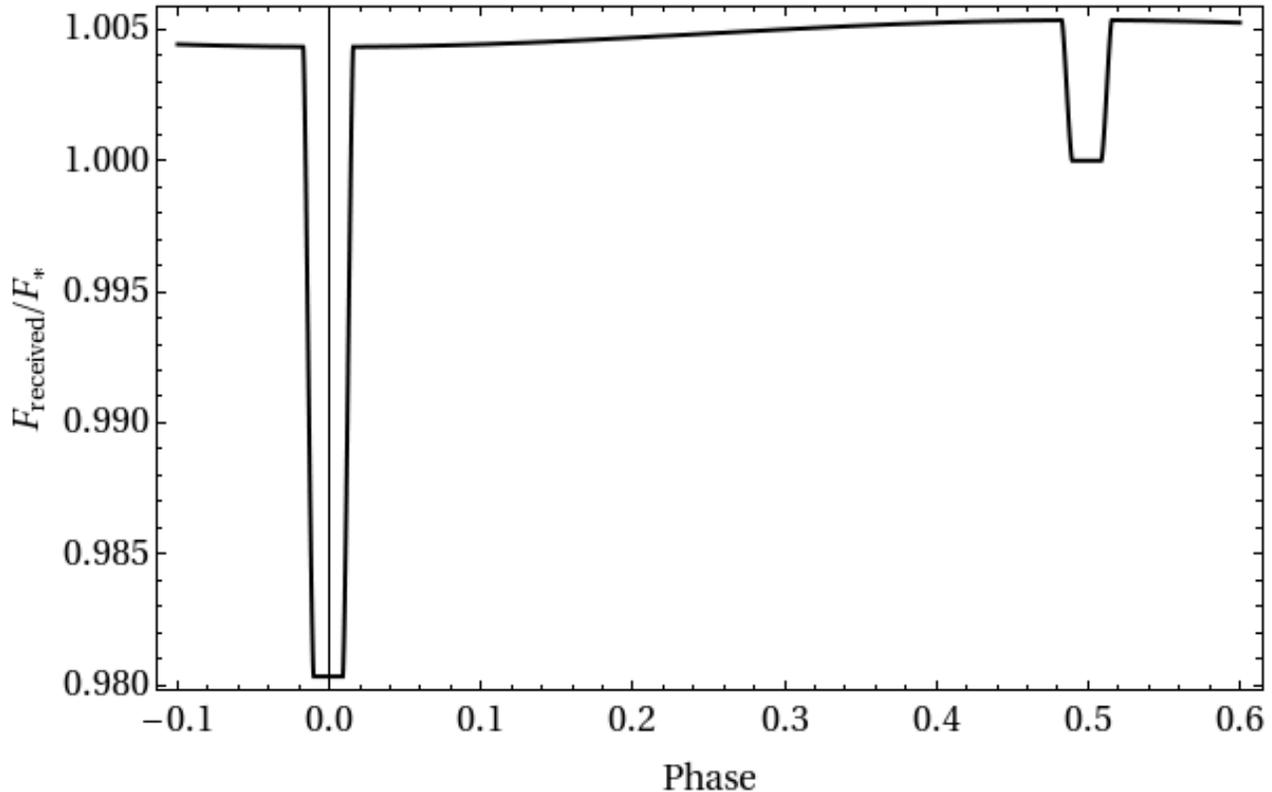}
\caption{\emph{Predicted phase curve of a hot-Jupiter with a hot dayside and cooler nightside.  Note how all fluxes are normalized to the star-only flux.}} \label{fig:fig1}
\end{center}
\end{figure*}

A phase curve time series is typically normalized to $F_*$, as shown in figure 1, which is in contrast to a normal transit measurement, which is normalized to $F_* + F_{\textrm{P,night}}$ i.e. the local out-of-transit baseline (see figure 2a for illustration).  Therefore, for a phase curve, the stellar normalized flux immediately before and after the transit event is equal to $B_{\mathrm{night}}$, as shown in figure 2b. It is possible that shifted hot spots on the planetary surface could cause an inequality between the pre and post transit baselines, but in practice the net effect of nightside pollution is very well accounted for by averaging over this time range.

\begin{figure*}
\begin{center}
\includegraphics[height=22.5 cm]{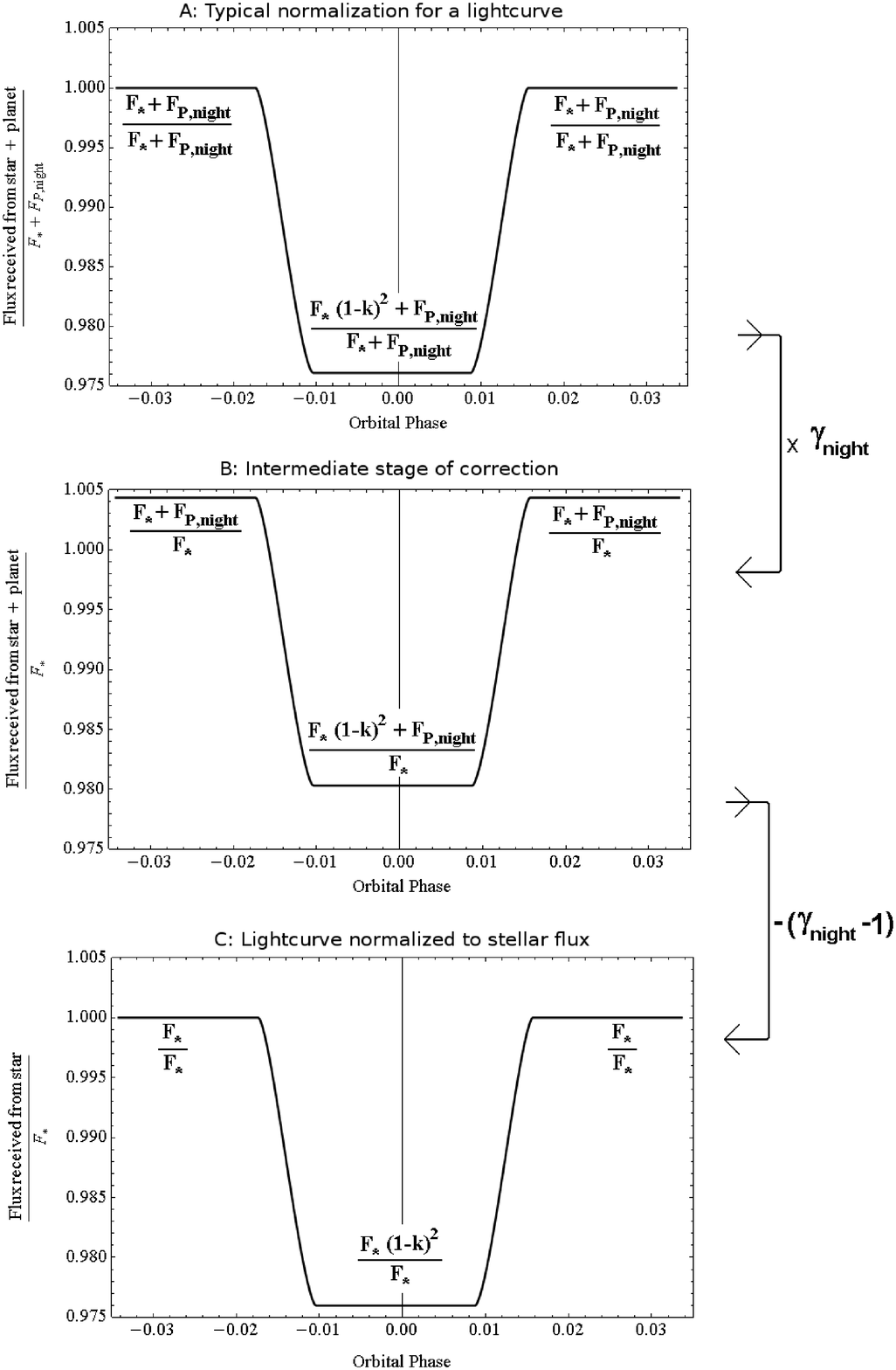}
\caption{\emph{Illustration of the three-stages involved in the corrective procedure, to compensate for the effects of nightside pollution.}} \label{fig:fig2}
\end{center}
\end{figure*}

In order to correct a primary transit lightcurve, we need to modify the normalization.  In figure 2, we show the two-step transformation which can achieve this.  We consider initially normalizing a lightcurve using the local baseline as usual for such measurements, as shown in figure 2a.  After this, the two-step correction may be performed, provided the observer has knowledge of $B_{\mathrm{night}}$.  The whole process may be summarized by the following (also summarized in figure 2):

\begin{enumerate}
\item[1.] Normalize fluxes to local out-of-transit baseline, as usual.
\item[2.] Multiply all flux values by $B_{\mathrm{night}}$.
\item[3.] Subtract $(B_{\mathrm{night}}-1)$ from all data points.
\end{enumerate}

In practice, these steps are incorporated into the lightcurve fitting algorithm directly. In the case of using Monte Carlo based routines for error estimation, $B_{\mathrm{night}}$ may be floated around its best-fit value and corresponding uncertainty. An example of this method is shown in \S5.1 and \S5.2 for the planet HD 189733b.  Defining $I_{j,\mathrm{uncorr}}$ as the locally normalized flux measurement of the $j^{\mathrm{th}}$ data point, we may explicitly write down the corrected data point as:

\begin{equation}
I_{j,\mathrm{corr}} = B_{\mathrm{night}} I_{j,\mathrm{uncorr}} - (B_{\mathrm{night}} - 1)
\end{equation}

We also briefly mention here that the transformation operations on the lightcurve time series will not only change the transit depth but also provides a more physical transit signal and thus we expect a slightly lower $\chi^2$ in the final fitting, as indeed is seen later in tables 1 and 2.

One caveat with the described method is the possible presence of ellipsoidal variations of the star, which would mix the phase curve signature. For example, \citet{wel10} detected ellipsoidal variations in HAT-P-7. Such signals peak at orbital phases of 0.25 and 0.75, whereas a phase curve should peak close to orbital phase 0.5, but can be offset by a small factor due to hot spots. \citet{wel10} provide a detailed discussion of modelling both signals and although ellipsoidal variations complicate the analysis, they certainly do not undermine it.

\subsection{Semi-empirical method}

To correct for the effect of nightside pollution, in an accurate way, we have proposed using phase curve information to obtain $B_{\mathrm{night}}$, which requires many hours of telescope time.  Further, the phase curve should be obtained at every wavelength simultaneously and for every epoch\footnote{Each epoch should be done in case temporal variability exists in the system} one wishes to measure the transit event at, in order to be sure of a completely reliable correction.  We label this resource-intensive method of correcting for the effect as the `empirical method'.  However, we appreciate that obtaining phase curves at every wavelength for each epoch is somewhat unrealistic and propose a `semi-empirical method' of achieving the same goal with far fewer resources.

We propose that observations of the phase curve are made at several wavelengths in the infrared; for example the IRAC and MIPS wavelengths of \emph{Spitzer} are very suitable but most of these channels are unfortunately no longer available.  With only a few measurements of the nightside flux, a prudent approach would be to assume the star and planet behave as blackbody emitters as a first approximation and extrapolate the emission to other wavelengths which are missing phase curve observations (an example of this will be provided in \S7).  Should higher resolution spectra of the nightside emission become available in the future, we may construct of a more sophisticated model as appropriate.  Transit observations at different wavelengths may then interpolate/extrapolate the model template to estimate the magnitude of the effect and apply the required correction at any wavelength. This allows us to estimate the blending factor at all wavelengths and times.

Long-term monitoring of the planet may also be necessary in order to ascertain the presence or absence of temporal variability in the system.  Ideally, this may be achieved by obtaining phase curves of the exoplanet to measure the nightside flux at regular times.  More practically, it could be done by measuring only the secondary eclipse, which is the dayside of the exoplanet, at regular times (for example \citet{ago09}).  Any large changes in the nightside flux are likely to be correlated to large changes in the dayside flux too, assuming a constant energy budget for the planet.  This second approach would reduce the demands on telescope time by an order of magnitude or more.

\subsection{Non-empirical method}

The final method we propose here is the least accurate but requires the fewest resources to implement.  If a planet has recently been discovered, no phase curves or even secondary eclipses may have been obtained yet.  Concordantly, the only avenue available is to estimate the temperature of the nightside through either a simple analytic estimation or a dynamical model of the atmosphere, although the latter may be excessive given the absence of any observational constraints.  We illustrate here how the temperature may be quickly estimated in such a case.

The nightside temperature may be estimated by assuming the dominant source of heating is from the incident stellar flux.  In this case, the only unknown factors affecting the nightside temperature are the re-distribution of heat factor, $f$ and the Bond albedo of the planet, $A_B$.  First-order estimations of these values can be made based upon empirical upper limits and measurements of other planets and atmospheric models.  The following expression may be used as a first-order estimate of the planet's brightness temperature:

\begin{equation}
T_{P,\mathrm{hemisphere}}(\lambda) = \Bigg(T_*^4 \frac{R_*^2}{4 a_P^2} [f (1-A_B)] + \frac{L_{\mathrm{int}}}{4 \pi R_P^2 \sigma_B \epsilon}\Bigg)^{1/4}
\end{equation}

Where $T_*$ is the effective temperature of the host star, $R_*$ is the stellar radius, $a_P$ is the orbital semi-major axis of the planet, $f$ is the distribution of energy to the hemisphere in question, $L_{\mathrm{int}}$ is the luminosity of the planet from internal heat generation (e.g. tidal heating, radioactivity), $\sigma_B$ is the Stefan-Boltzmann constant and $R_P$ is the radius of the planet.

For this calculation, $L_{\mathrm{int}}$ is generally assumed to be zero unless large tidal forces are expected as a result of a highly eccentric orbit, for example.  All of the other quantities are typically measured except for $f$ and $A_B$.  Choices for these values may come from atmospheric models or experience with other exoplanets.

\section{Comparison to other effects}
\subsection{Starspots}

Starspots have been observed within the transit events in several cases; e.g. \citet{pon07}, \citet{dit09}. They typically have been observed to have a radius of a few Earth radii and are estimated to have temperatures from 100K to 1000K cooler than the rest of the stellar surface. When a planet passes over a starspot, it results in an increase in relative flux within the transit signal which is easily identified. If one assumes that the only starspot is the starspot which has been crossed, then accounting for the effect is quite trivial and may be incorporated in the lightcurve modelling.

What is much more troublesome are out-of-transit starspots for which we have no direct evidence. The presence of out-of-transit starspots will cause the transit depth to appear larger, in general. As a typical example, \citet{cze09} estimate that the effect can cause underestimations of the planetary radius by a fraction of $\sim 3$\% for CoRoT-2b, which is a 1.6\% change in the transit depth. This effect is larger than the nightside pollution effect by a factor of 5-10.

However, these effects can only be present for spotty stars whereas nightside pollution simply requires a hot planet. Also, the spot coverage of a stellar surface varies periodically giving rise to an ultimately regular pattern which may therefore be corrected for. In contrast, the nightside pollution effect is not periodic, it is a constant offset in a single direction. Further, prior information such as a phase curve or a dayside eclipse places strong constraints on $B_{\mathrm{night}}$ and therefore the estimation of this parameter is not an issue. Therefore, even for spotty stars, no-one would propose ignoring the effects of blending induced by a nearby companion star and so it can be seen that negating the self-blend of the planet would also be folly.

\subsection{Temporal variability of $F_P$ and $F_*$}

If the stellar flux or the nightside emission of the planet experiences temporal variations, then we would expect $d_{\mathrm{obs}}$ to also change over time.  We will here estimate the magnitude of this effect.  The expected changes in nightside emission has not been studied in as much detail as that for the dayside, but we expect the magnitude of variations to be very similar. \citet{rau07} have used shallow-layer circulation models to estimate variations in the dayside emission at the 1\%-10\% level.  This is consistent with the observations of HD 189733b's dayside flux by \citet{ago09} who measure variations in the dayside below 10\%. Typical stellar flux changes are at the 1\% level and so the ratio $F_P/F_*$ is more likely to vary due to the planet than the star.

For the case of HD 189733b, we later show that $B_{\mathrm{night}} = 1.002571$ at 8.0$\mu$m.  If $F_{\mathrm{night}}$ increased by $\pm 10$\%, this would correspond to $B_{\mathrm{night}} = 1.002828$ causing the transit depth to vary by 0.0006\%.  This would be around an order of magnitude below \emph{Spitzer}'s sensitivity but could be potentially close to a 1-$\sigma$ effect for JWST.  Nevertheless, the effect is sufficiently small that it is unlikely to be significant in most cases.

\subsection{Limb darkening}

The nightside pollution effect is generally only relevant for hot-Jupiters at infrared wavelengths. As we move towards the 10-30 $\mu$m wavelength range, the effects of limb darkening become negligible. The curvature of the transit trough is essentially flat. However, the limb of the star will possess a more complicated profile (\citet{oro00}; \citet{jef06}) and this could potentially introduce errors into the fitting procedure. It is generally prudent to include even the very weak limb darkening effects when modelling such transits.

As a result, equations (6) and (7) should not be used to attempt to correct lightcurves which were fitted without nightside pollution. They do, however, offer a useful approximate quantification as to the direction and magnitude of any errors. A comparison between the predictions of (6) and (7) and the exact limb-darkened nightside-polluted lightcurve fits is given later in \S5.1 for the example of HD 189733. A discussion of this method required to produce this exact modelling is given in \S3.1.

\subsection{Significance at visible wavelengths}

We briefly consider the value of including the nightside pollution effect at visible wavelengths, in particular for the \emph{Kepler Mission}. \citet{bor09} recently reported visible-wavelength photometry for HAT-P-7b which exhibits a combination of ellipsoidal variations and a phase curve \citep{wel10}, as well as a secondary eclipse of depth $(130 \pm 11)$ppm. HAT-P-7b is one of the very hottest transiting exoplanets discovered, so it offers a useful upper-limit example. For the purposes of nightside pollution, the maximum possible effect would occur if eclipse was both due to thermal emission alone and efficient day-night circulation. In this hypothetical scenario which maximizes the nightside pollution, we would have $B_{\mathrm{night}} = 1.00013$. In the case of HAT-P-7b, the transit depth was reported to be $d_{\mathrm{obs}} = (6056 \pm 47)$ppm by \citet{wel10}, implying that the geometric transit depth is larger by $0.79$ppm, or $0.017$-$\sigma$. Therefore, as expected, visible wavelength transits will not, in general, be significantly affected by nightside pollution for even \emph{Kepler} photometry.

\section{Applied Example - HD 189733b}
\subsection{\emph{Spitzer} IRAC 8.0$\mu$m measurement}

We will here provide an example of the empirical method of compensating for nightside pollution. We used the corrected data of HD 189733's phase curve at 8.0$\mu$m, as taken by \citet{knu07} (obtained through personal communication).  We applied a median-stack smoothing function to the lightcurve with a one-minute window in order to identify the eclipse contact points.  We find the flux of the star by taking the mean of fluxes between the $2^{\mathrm{nd}}$ and $3^{\mathrm{rd}}$ contact points during the secondary eclipse, weighting each point by the reported error.  The standard deviation within this region is divided by the square root of the number of data points to give us the error on the mean.  All fluxes are then divided by the derived stellar flux and the error on each flux stamp is propagated through, incorporating the error on the stellar flux estimate.

In order to determine the nightside flux, which is not the same as the minimum flux, we adopt a baseline defined as 30 minutes before $1^{\mathrm{st}}$ contact and 30 minutes after $4^{\mathrm{th}}$ contact and find a mean of $B_{\mathrm{night}} = 1.002571 \pm 0.000048$.  The average rms in this baseline is 0.65 mmag/minute.  If it were possible for the nightside of the planet to induce a secondary eclipse, as the dayside does, we would measure a secondary eclipse nightside depth of $(0.256 \pm 0.023)$\%, whereas \citet{cha08} report a dayside secondary eclipse depth of $(0.391 \pm 0.022)$\%.

As discussed in \S3.1, ellipsoidal variations can also be responsible for out-of-transit flux variations and can be potentially confused with the phase curve. For HD 189733, we use equation (1) of \citet{pfahl08} to estimate an ellisoidal variation amplitude of $2.2$ppm. Given that the phase curve exhibits a variation $1350$ppm amplitude, ellipsoidal variations can be neglected for the rest of this analysis.

We now produce two fits of the lightcurve: 1) no blending factor 2) blending factor $B_{\mathrm{night}}$ included.  Each lightcurve is fitted independently assuming a fixed period of $P = 2.2185733$ days and zero orbital eccentricity.  The results of the fits are displayed in table 1.

For the fitting, we use a Markov Chain Monte Carlo (MCMC) algorithm which employs the geometric model of \citet{kip08}, the limb darkening model of \citet{man02} and utilizes the lightcurve fitting parameter set described in \citet{kip10}: $t_C$, $b^2$, $\Upsilon/R_*$, $p^2$ and $OOT$.  We use 125,000 trials with the first 25,000 discarded for burn-in. Employing the code of I. Ribas, a \citet{kur06} style atmosphere is used to interpolate the four non-linear limb darkening coefficients \citep{cla00}, following the same methodology of \citet{bea10}, giving us $c_1 = 0.2790207$, $c_2 = -0.1506885$, $c_3 = 0.0779481$ and $c_4 = -0.0087653$.  We use the same local baseline as defined earlier, constituting 22382 data points and assume a circular orbit.  At this stage, no outliers have yet been rejected but we proceed to fit the unbinned lightcurve.  We take the best-fit lightcurve and subtract it from the data to obtain the residues and then look for outlier points.  We use the median-absolute-deviation (MAD) (\citet{gau16}) to provide a robust estimate of the standard deviation of the data, as this parameter is highly resistant to outliers, and find $\textrm{MAD} = 2.92062 \times 10^{-3}$.  Since there are $22382$ points, then the maximum expectant departure from a normal distribution is 4.08 standard deviations.  Any points above this level are rejected, where our definition of the standard deviation comes from the MAD value multiplied by 1.4826, as appropriate for a normal distribution\footnote{Although strictly a Poisson distribution, for $22382$ data points, the distribution is very well approximated by a Gaussian}.  This procedure rejects any points with a residual deviation greater or equal to 0.0176749, corresponding to 10 points\footnote{We note that the data has already been cleaned of outlier measurements, which is why the MAD rejection criteria only identifies 10 outliers from $22382$ points}.

\begin{table*}
\caption{\emph{Best-fit transit parameters for the HD 189733b 8.0$\mu$m primary transit lightcurve; data obtained by \citet{knu07}.  Fits performed for the case of 1) typical normalization the local baseline 2) correction for the effects of nightside pollution.  The number of data points is 22372.}} 
\centering 
\begin{tabular}{l c c c c c c} 
\hline\hline 
Method & Depth, $p^2$,\% & $T$/s & $b$ & $a/R_*$ & $i$/$^{\circ}$ & $\chi^2$ \\ [0.5ex] 
\hline 
(1) Local baseline & 2.3824 & 5127.45 & 0.66264 & 8.9121 & 85.7360 & 22412.4691 \\
(2) Nightside correction & 2.3884 & 5127.55 & 0.66204 & 8.9183 & 85.7428 & 22412.4605 \\
Uncertainity & 0.0061 & 8.1 & 0.0061 & 0.050 & 0.054 & - \\
\hline
(2) - (1) & +0.0060 & +0.10 & -0.00060 & +0.0062 & +0.0068 & -0.0086  \\ [1ex]
\hline\hline 
\end{tabular}
\end{table*}

The new lightcurve is then refitted in the normal way and we present the best-fit value in table 1.  We find the uncorrected lightcurve has a depth of  $d_{\mathrm{obs}} = 2.3824 \pm 0.0061$\%.  For comparison, we note that \citet{knu09} report the fitted 8.0$\mu$m depth to be $2.387 \pm 0.006$\%, which is consistent with our value.  Applying the correction due to nightside pollution, we find the geometric transit depth to be $d_{\mathrm{geo}} = 2.3884 \pm 0.0061$\% meaning that the depth has increased by $0.006$\% corresponding to $1$-$\sigma$.  From this example, it is clear that negating an effect which systematically biases transit depth measurements by $\sim1$-$\sigma$ would be imprudent.

Using equation (7), to first-order in $(B-1)$, we estimate that the impact parameter should be overestimated by 0.00069. Our fits reveal a very similar figure of 0.00060.  Similarly for $a/R_*$ we predict an underestimation of 0.0034 whereas the lightcurve fit finds 0.0062. As expected, the effect of a blend is less pronounced on the other parameters.

Based on the difference in collecting area, it is expected JWST will achieve a precision $\sim 6.6$ times greater than \emph{Spitzer}, suggesting this nightside pollution effect will become significant at the $\sim 5$-$10$-sigma level for future infrared transit observations of hot-Jupiters.  Additionally, the binning of multiple \emph{Spitzer} transits would raise the significance of the effect. For example, \citet{ago09} reported seven 8.0$\mu$m transits of HD 189733b which, if globally fitted would inrease the significance of the nightside pollution effect to 2.6-$\sigma$. Such a large effect cannot be justifiably negated.

\subsection{\emph{Spitzer} MIPS 24$\mu$m}

\citet{knu09} measured the phase curve of HD 189733b with the MIPS instrument onboard \emph{Spitzer} about a year after the observations of the 8.0$\mu$m phase curve for the same system.  Using the original normalized-to-stellar-flux unbinned data (personal correspondence with H. Knutson), we took the mean of data points $\simeq 1$ hour either side of the transit event, which exhibit an rms of 2.1 mmag/minute.  We combined to two baseline estimates to calculate a nightside relative flux of $B_{\mathrm{night}} = 1.00438 \pm 0.00025$.

\citet{knu09} reported a 24$\mu$m transit depth of $2.396 \pm 0.027$\% and our own re-analysis of the data yields $d_{\mathrm{obs}} = 2.398 \pm 0.019$\%, where the fit has been performed using the same methodology as for 8.0$\mu$m, except we assume no limb darkening at 24$\mu$m. As before, we apply the correction due to nightside pollution and estimate a new 24$\mu$m transit depth of $d_{\mathrm{geo}} = 2.409 \pm 0.020$\%, which increases the depth by $\sim 0.5$-$\sigma$.  Despite the absolute effect being larger than that at 8.0$\mu$m, the difference is fewer standard deviations away due to the much poorer signal-to-noise of the transit event itself.

\begin{table*}
\caption{\emph{Best-fit transit parameters of the HD 189733b 24.0$\mu$m primary transit lightcurve; data obtained by \citet{knu09}.  Fits performed for the case of 1) typical normalization the local baseline 2) correction for the effects of nightside pollution.  Number of data points is 1198.}} 
\centering 
\begin{tabular}{l c c c c c c} 
\hline\hline 
Method & Depth, $p^2$,\% & $T$/s & $b$ & $a/R_*$ & $i$/$^{\circ}$ & $\chi^2$ \\ [0.5ex] 
\hline 
(1) Local baseline & 2.3980 & 5072.527 & 0.61425 & 9.492 & 86.290 & 1260.9125 \\
(2) Nightside correction & 2.4085 & 5072.505 & 0.6131 & 9.502 & 86.300 & 1260.9092 \\
Uncertainity & 0.019 & 19.7 & 0.003 & 0.033 & 0.022 & - \\
\hline
(2) - (1) & +0.011 & -0.022 & -0.0012 & +0.010 & +0.010 & -0.0033 \\ [1ex]
\hline\hline 
\end{tabular}
\end{table*}

\section{Pollution of the transmission spectrum}

In the standard theory of transmission spectroscopy, planetary nightside emission is assumed to be negligible and thus disregarded (e.g. as explicitly stated in the foundational theory of \citet{brown01}).  However, we have shown here that the effect noticeably changes the transit depth for high quality photometry.  Essentially, we posit that the `traditional' transmission spectrum is in fact a combination of transmission through the terminator and the self-blending caused by emission from the nightside. 

One subtle point is that the nightside pollution effect is not something which can be accounted for in the modelling of the transmission spectrum. It is generally useful to think of the nightside pollution effect as an astrophysical blend which happens to be related to the planetary properties. A transmission spectrum is typically found by fitting a transit lightcurve at multiple wavelengths and then fitting a spectrum through the retrieved transit depths which models the planetary atmosphere. These routines usually make use of radiative transfer, chemical equilibrium, molecular line lists, etc to estimate the opacity of the atmosphere at each wavelength. However, attempting to increase the sophistication of these models would not accurately account for the self-blending scenario. Recall that each transit depth is obtained by fitting an eclipse model through the lightcurve time series. Critically, it is at this stage where blending needs to be accounted for. The transit signal plus blend should be modelled as such from the outset due to subtle and quite intricate inter-dependencies between $b$, $a/R_*$, $p^2$ and the limb darkening. Thus it can be seen that attempting to incorporate the effects later on is far more challenging and completely unneccessary than simply fitting each transit lightcurve with a physically accurate model in the first place.

In this section, we will estimate how different an exoplanet's transmission spectrum would appear with and without nightside pollution. In order to evaluate the magnitude of the effect, we will here consider a planet of similar type to HD 189733b.  It is important to stress that the effects of nightside pollution will vary from case to case and the example we give here is indeed just for one example which is somewhat typical for an observed hot-Jupiter.  Therefore the results here are only for a hypothetical, but typical, example.

The real question we need to answer is how much does a planetary transmission spectra change due to nightside self-blending?  We therefore need to generate two versions of the planetary transmission spectra, one including (figure 3a) and one excluding the effects of nightside pollution (figure 3c), and then take the difference between the two (figure 3d).

For a description of the models used to generate the spectra, details may be found in \citet{tin07a} and \citet{tin07b} for the transmission spectrum and \citet{swa09} for the emission.  Planet and star properties are set to be that of the HD 189733 system.  The model contains water, carbon dioxide and methane to provide us with the effects of molecular species on nightside pollution.  No carbon monoxide or hazes/clouds are included in our example.  We note that the transmission and emission models are good fits to the current available spectroscopy/photometry data of HD 189733b in the NIR/MIR \citep{swa09}.  The effects of water absorption are quantified with the BT2 water line list \citep{bar06}, which characterises water absorption at the range of temperatures probed in HD 189733b. Methane was simulated by using a combination of HITRAN 2008 \citep{rot05} and PNNL data-lists.  Carbon dioxide absorption coefficients were estimated with a combination of HITEMP and CDSD-1000 (Carbon Dioxide Spectroscopic Databank version for high temperature applications; Tashkun \& Perevalov (2008)). The continuum was computed using $H_2-H_2$ absorption data \citep{bor01}.

Generated spectra are always plotted in terms of the quantities determined with the lowest measurement uncertainty, namely $(R_P/R_*)^2$ and $(F_P/F_*)$, for the primary and secondary eclipses respectively.  Transmission spectra which are plotted in units of $R_P$ will cause the measurement uncertainties to be much larger since the error on $R_*$ must necessarily be propagated in such a recipe.  In fact, the measurement uncertainties on a spectra plotted in units of $R_P$ will be dominated by the error on $R_*$ since this property is typically determined to much lower precision.  Consequently, statistically significant molecular features would be overwhelmed by the artificially large error bars.

Using the model described above, we first compute the transit depth from transmission absorption effects only (i.e. excluding nightside pollution) as visible in figure 3a.  We then generate the dayside emission spectra for the same planet and make the assumption that the dayside and nightside emission spectra are identical (figure 3b).  This assumption is unlikely to be true for the exact case of HD 189733b and really constitutes an upper limit, but we again stress that we are here only considering a planet similar to that of HD 189733b and thus we are free to make this assumption for our hypothetical example.

\begin{figure*}
\begin{center}
\includegraphics[width= 16.8 cm]{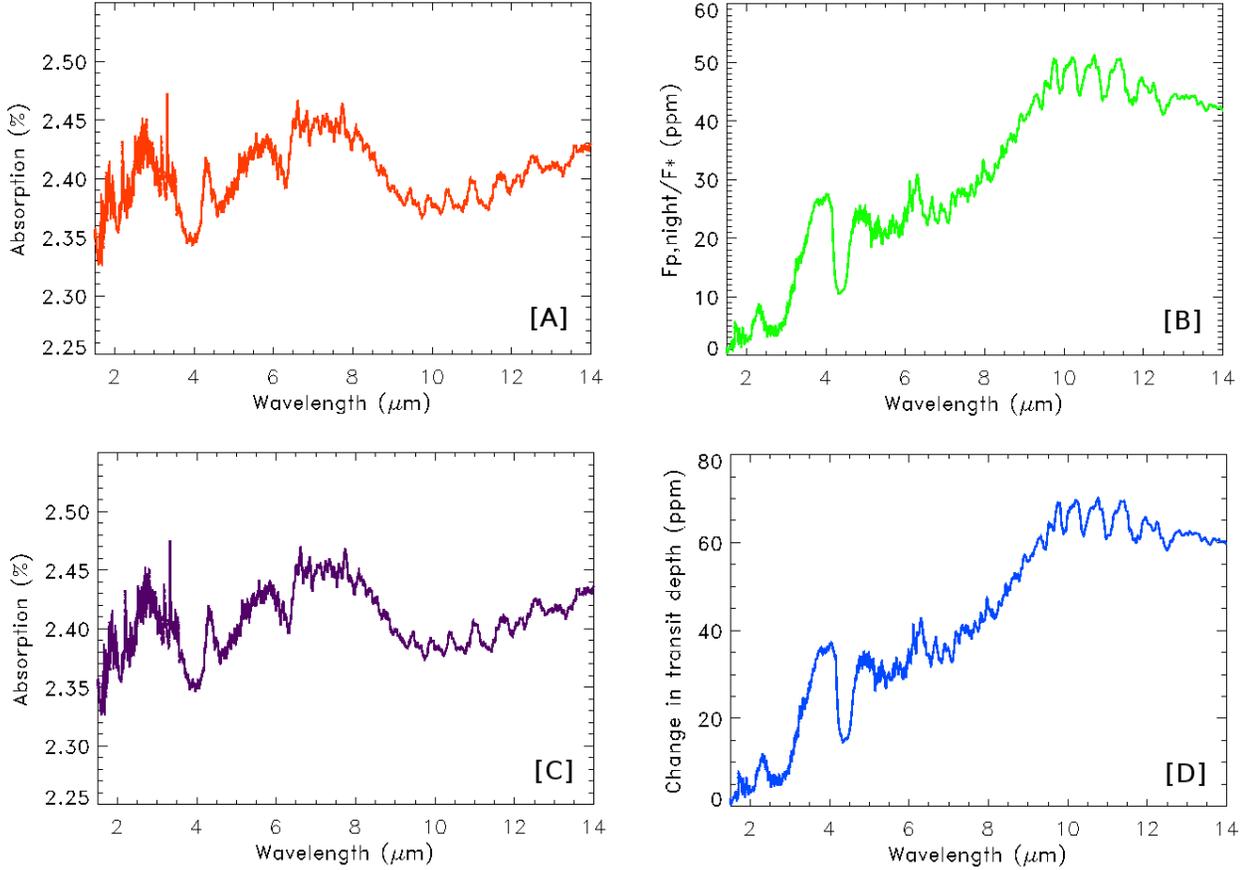}
\caption{\emph{Top left:} \textbf{A}- Transmission spectrum of a hypothetical exoplanet similar to HD 189733b, generated considering the transmission through the terminator only.  \emph{Top right:} \textbf{B}- Emission spectrum from the nightside of the hypothetical planet.  \emph{Bottom left:} \textbf{C}- Transmission spectrum of the planet incorporating the pollution of the nightside emission.  \emph{Bottom right:} \textbf{D}- Residual between two transmission spectra.  We conclude that not accounting for nightside emission would result in a 60-80ppm error in the transit depth.} \label{fig:fig3}
\end{center}
\end{figure*}

\begin{figure}
\begin{center}
\includegraphics[width= 8.4 cm]{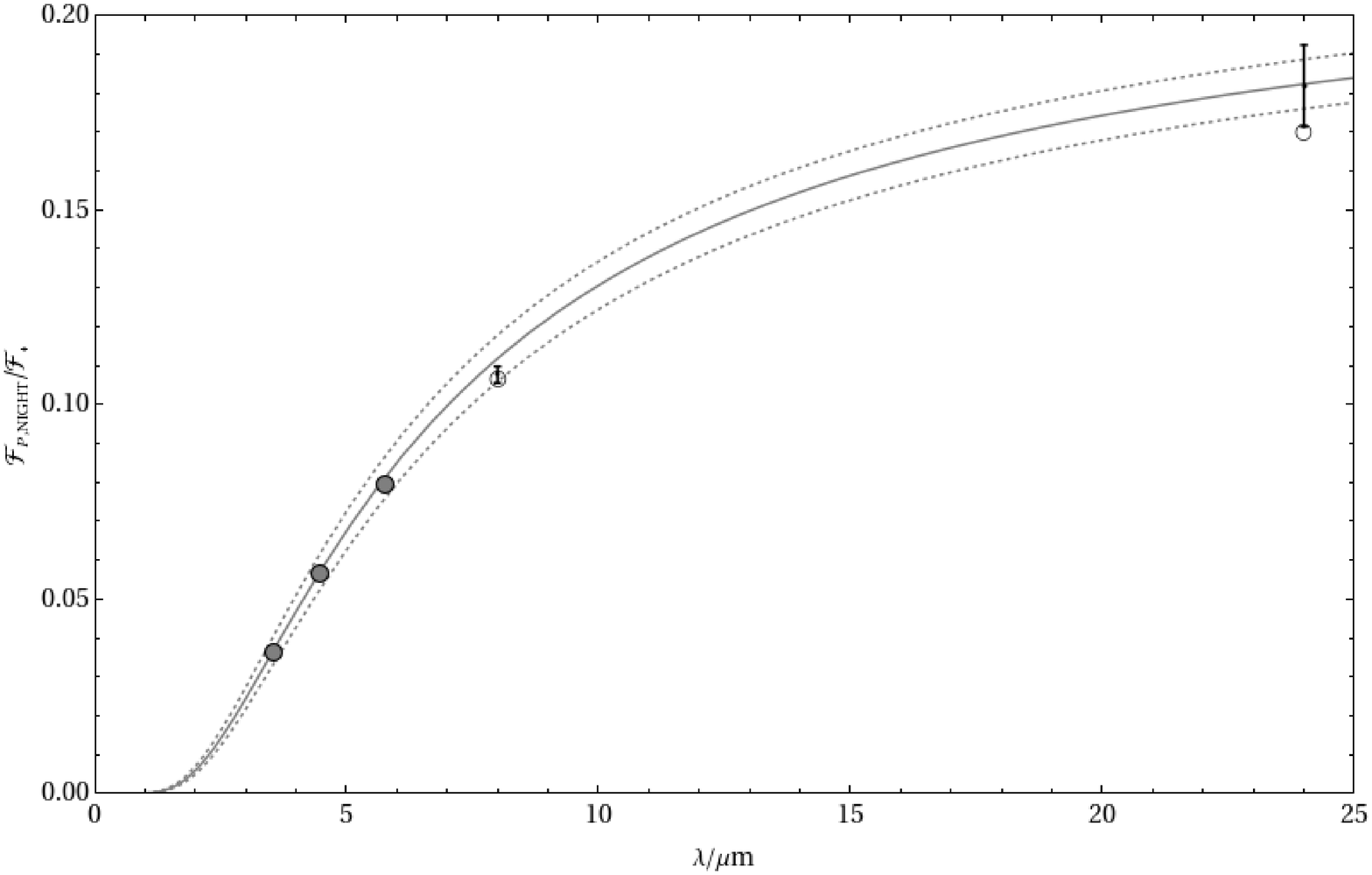}
\caption{\emph{Flux per unit area (i.e. not total flux) of the planetary \textbf{nightside} emission divided by that of the star, plotted as a function of wavelength.  Using the two measurements at 8.0$\mu$m and 24.0$\mu$m (black dots with error bars), we fit a blackbody curve through the points (gray lines) with a nightside planetary temperature of $T = 1148 \pm 32$K.  Open circles represent the integrated blackbody function across the instrument bandpasses.  Filled circles represent the same but extrapolated to the other IRAC wavelengths, which allow us to conclude the nightside effect will be much less at lower wavelengths.}} \label{fig:fig4}
\end{center}
\end{figure}

We combine the nightside emission and transmission spectra to produce a transmission spectra which includes the effects of the nightside, as seen in figure 3c. We then take the difference between the corrected spectrum and the one which excludes nightside emission. The resultant residual spectrum is plotted in figure 3d.  The residual spectrum reveals nightside emission affects the transmission spectra at the level of $6 \times 10^{-5}$ above 10 microns and very closely matches the behavior of the emission features, as expected.  As we saw earlier, the magnitude of the effect is equal to the typical measurement uncertainty for a target like HD 189733b with \emph{Spitzer}. This supports our hypothesis that the nightside pollution effect is a $\sim$1-$\sigma$ effect for 8.0$\mu$m \emph{Spitzer} photometry.

\section{Extrapolating the Nightside Correction for HD 189733b}

Only two measurements exist for the nightside flux for HD 189733b (or indeed any other exoplanet) at 8.0$\mu$m and 24$\mu$m, but several other primary transit lightcurves exist in the mid-infrared wavelengths.  \citet{bea08} presented 3.6$\mu$m and 5.8$\mu$m measurements and \citet{des09} obtained photometry at 4.5$\mu$m and 8.0$\mu$m.  We will here estimate the nightside effect on the 3.6$\mu$m, 4.5$\mu$m and 5.8$\mu$m channels.  Currently, only two data points exist and this does not warrant us modelling the nightside emission spectrum in any more complexity than that of a blackbody, as a first order approximation.

We first assume that the star is blackbody emitter with $T_* = 5040 \pm 50$K \citep{tor08}.  At this point is advantageous to consider only the emission per unit area from both the planet and the star, in order to avoid the effects of transmission through the planetary atmosphere.  We therefore define the flux per unit area of each object using:

\begin{align}
\mathcal{F}_{P,\mathrm{night}} &= F_{P,\mathrm{night}}/(\pi R_P^2) \\
\mathcal{F}_* &= F_*/(\pi R_*^2) \\
\mathcal{R}_{\mathrm{night}} &= \frac{\mathcal{F}_{P,\mathrm{night}}}{\mathcal{F}_*} = \frac{F_{P,\mathrm{night}}}{F_*} d_{\mathrm{geo}}
\end{align}

We then convert the $B_{\mathrm{night}}$ measurements for 8.0$\mu$m and 24$\mu$m into $\mathcal{R}_{\mathrm{night}}$ by using the two values of the corrected ratio-of-radii we showed in tables 1 and 2 and propagating the uncertainties.  Using equations (6) \& (7) and the relevant \emph{Spitzer} bandpass response functions, we are able to estimate $\mathcal{R}_{\mathrm{night}}$ for any given value of $T_{P,\mathrm{night}}$.  We allow this temperature to vary from 700K to 1700K in 1K steps and numerically integrate the bandpasses to find $\chi^2$ at each temperature, which we define as:

\begin{align}
\chi^2 &= \Bigg(\frac{\mathcal{F}_{\mathrm{obs}} - \mathcal{F}_{\mathrm{calc}}}{\Delta(\mathcal{F}_{\mathrm{obs}})}\Bigg)^2\Bigg|_{8.0\mu\mathrm{m}} + \Bigg(\frac{\mathcal{F}_{\mathrm{obs}} - \mathcal{F}_{\mathrm{calc}}}{\Delta(\mathcal{F}_{\mathrm{obs}})}\Bigg)^2\Bigg|_{24\mu\mathrm{m}} \nonumber \\
\qquad& + \Bigg(\frac{T_{*,obs} - T_{*,calc}}{\Delta(T_*)}\Bigg)^2
\end{align}

Where we additionally define $T_{*,\mathrm{calc}}$ as the temperature used in the integration and $T_{*,obs}$ as being equal to the value determined by \citet{tor08}.  Note that we do not fit for $T_*$ but do allow the value to float in order to correctly estimate the uncertainity of $T_{P,\mathrm{night}}$.  The final analysis reveals a best-fit planetary nightside temperature of $T_{P,\mathrm{night}} = 1148 \pm 32$K with $\chi^2 = 1.12$, suggesting the blackbody model gives a satisfactory fit for these two measurements.  We note that not accounting for the response function of the instruments would yield a erroneous result of $T_{P,\mathrm{night}} = 1120$K.

Using our derived planetary temperature, we may now use the blackbody function to extrapolate $\mathcal{R}_{\mathrm{night}}$ to other wavelengths and thus the nightside corrected transit depths for 3.6$\mu$m, 4.5$\mu$ and 5.8$\mu$m.  The geometric transit depth will be given by equation (14) and the results of our analysis our summarized in table 3 and figure 4.

\begin{equation}
d_{\mathrm{geo}} = \frac{d_{\mathrm{obs}}}{1-\mathcal{R}_{\mathrm{night}}d_{\mathrm{obs}}}
\end{equation}

\begin{table*}
\caption{\emph{Using a fitted blackbody function for the nightside emission of HD 189733b, we calculate the nightside corrections to \emph{Spitzer} channels for which no phase curve information currently exists. Values with a $\dagger$ superscript cannot have their uncertainties estimated since they are extrapolated parameters.  8.0$\mu$m data comes from \citet{knu07}, 24.0$\mu$m from \citet{knu09}, 3.6$\mu$m \& 5.8$\mu$m from \citet{bea08} and 4.5$\mu$m from \citet{des09}.}} 
\centering 
\begin{tabular}{l c c c c} 
\hline\hline 
Channel & Observed depth $d_{\mathrm{obs}}$,\% & $\mathcal{R}_{\mathrm{night}}$ & Corrected depth $d_{\mathrm{geo}}$,\% & $\frac{d_{\mathrm{geo}} - d_{\mathrm{obs}}}{\sigma_{d}}$ \\ [0.5ex] 
\hline
\emph{Measured} \\
\hline
8.0$\mu$m & $2.3824 \pm 0.0060$ & $0.1076 \pm 0.0020$ & $2.3884 \pm 0.0061$ & 1.0 \\
24.0$\mu$m & $2.398 \pm 0.019$ & $0.1818 \pm 0.0105$ & $2.4085 \pm 0.020$ & 0.5 \\ 
\hline 
\emph{Extrapolated} \\
\hline
3.6$\mu$m & $2.356 \pm 0.019$ & $0.0365^{\dagger}$ & $2.358 \pm 0.019^{\dagger}$ & $\lesssim 0.1$ \\
4.5$\mu$m & $2.424 \pm 0.010$ & $0.0570^{\dagger}$ & $2.427 \pm 0.010^{\dagger}$ & $\lesssim 0.4$ \\
5.8$\mu$m & $2.436 \pm 0.020$ & $0.0800^{\dagger}$ & $2.441 \pm 0.020^{\dagger}$ & $\lesssim 0.25$ \\[1ex]
\hline\hline 
\end{tabular}
\end{table*}

The maximal deviation occurs for 8.0$\mu$m and is less than 1-$\sigma$ for all other wavelengths.  Consequently, the deduction of which molecules are evident from the spectrum of HD 189733b will not significantly affected by the nightside effect, but derived abundances will change. In future work, we will re-process and re-interpret the HD 189733b spectrum including near-infrared measurements and the nightside effect, but a detailed re-processing of these data is outside of the scope of this theoretical paper.

\begin{figure}
\begin{center}
\includegraphics[width= 8.4 cm]{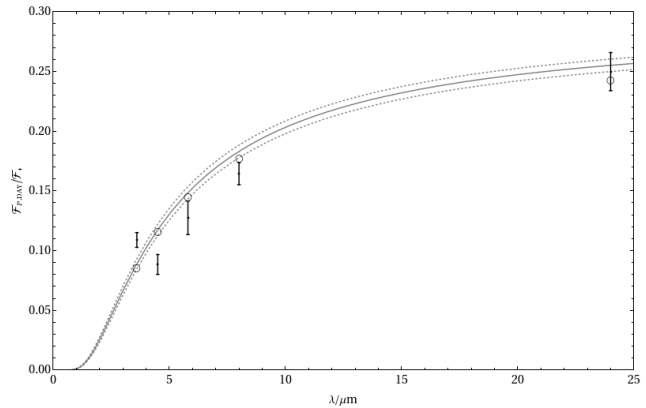}
\caption{\emph{Flux per unit area (i.e. not total flux) of the planetary \textbf{dayside} emission divided by that of the star, plotted as a function of wavelength.  Using the available measurements (black dots with error bars), we fit a blackbody curve through the points (gray lines) with a dayside planetary temperature of $T = 1490 \pm 25$K.  Open circles represent the integrated blackbody function across the instrument bandpasses.  The nightside and dayside temperatures suggest a redistribution factor of 0.53}} \label{fig:fig5}
\end{center}
\end{figure}

Combining the corrected geometric transit depths of table 3 with secondary eclipse measurements from the exoplanet literature, we may estimate the dayside brightness temperature in a similar manner, for completion.  We use the secondary eclipse depth measurements quoted by \citet{cha08} for 3.6$\mu$m, 4.5$\mu$m, 5.8$\mu$m, 8.0$\mu$m and 24$\mu$m.  Note, we we do not use the 16$\mu$m data point since no primary transit exists for this wavelength which we can use to estimate to $\mathcal{R}_{\mathrm{day}}$.  A $\chi^2$ minimization between 1000K and 2000K in 1K steps, integrating over each bandpass in every simulation, reveals a best-fit dayside temperature of $T_{P,\mathrm{day}} = (1490 \pm 25)$K with $\sqrt{\chi^2}|_{\mathrm{red}} = 1.45$ (for comparison $T_{P,\mathrm{night}} = (1148 \pm 32)$K).  In figure 5, we plot $\mathcal{F}_{P,\mathrm{day}}/\mathcal{F}_*$, in an analagous way as we did for the nightside in figure 4.

Assuming the internal heating of the planet is negligible and the planet is tidally locked, we may combine the two temperature estimates and employ equation (9) to obtain the redistribution factor of heat to the nightside, $f_{\mathrm{night}} = 0.521 \pm 0.050$.  \citet{bur08} predict the redistribution should be in the range $\sim 0.1$ to $\sim 0.4$ meaning our estimate may be difficult to explain theoretically.

\section{Conclusions}

We have shown that emission from the nightside of an exoplanet acts as a self-blend causing transit depths to be systematically erroneous, more specifically, underestimated.  This effect is largest for infrared measurements of hot-Jupiter systems and can cause changes in the transit depth at the level of $10^{-4}$, which is around the 1-$\sigma$ level per transit for \emph{Spitzer} measurements of bright stars.  Crucially, this is also the magnitude of the molecular features induced in the atmospheric transmission spectrum. The multiple observed \emph{Spitzer} transits binned together raises the significance of the effect to 2.6-$\sigma$ and for JWST we estimate the effect will be $5$-$10$-$\sigma$ significant per transit.

The effect may be compensated for in an accurate manner by using observations of phase curves of exoplanets to obtain the nightside emission flux directly and then using this value to correct transit depths.  Naturally, this requires observations of both the transit event and the phase curve in question to be at the same wavelength and ideally the same epoch to avoid problems associated with temporal variability.  A more practical approach may be to obtain phase curves at several wavelengths and use a model-fitted temperature profile to interpolate/extrapolate the required correction for other observations.

The effect has a strong spectral dependency and therefore should be incorporated in transit lightcurve fits before modelling the transmission spectrum, in order to obtain the most accurate interpretation possible.  This is particularly true when looking for absorption features in the infrared at the level of $10^{-4}$, which is common for many molecular species already detected (for example water vapour and methane, \citet{swa08}).

\section*{Acknowledgments}

We thank the anonymous referee for their helpful comments in revising this manuscript. D. M. K. has been supported by UCL, the Science Technology and Facilities Council (STFC) studentships, the Harvard-Smithsonian Center for Astrophysics and NASA grant NNX08AF23G.  Special thanks to H. Knutson et al. for kindly providing us with their photometry and I. Ribas for providing limb darkening coefficients.  We would like to thank S. Fossey, M. Swain, J. P. Beaulieu, G. Bakos, G. Vashisht, and A. Aylward for their support and discussions in preparing this manuscript.  G.T. is supported by UCL and the Royal Society.

\bsp

\label{lastpage}

\end{document}